\documentclass[
    ,final            
  ]
  {aipproc}

\layoutstyle{6x9}

\begin{document}
\title{GPD Physics With Polarized Muon Beams at COMPASS-II}
\author{Andrea Ferrero (on behalf of the COMPASS collaboration)}{address={CEA-Saclay, DSM/Irfu/SpHN, 91191 Gif-sur-Yvette (France)}}
\begin{abstract}
A major part of the future COMPASS program is dedicated to the investigation
of the nucleon structure through Deeply Virtual Compton Scattering (DVCS) and Deeply Virtual Meson Production (DVMP). COMPASS will measure DVCS and DVMP reactions with a high intensity
muon beam of 160 GeV and a 2.5 m-long liquid hydrogen target surrounded
by a new TOF system. The availability of muon beams with high energy
and opposite charge and polarization will allow to access the Compton
form factor related to the dominant GPD $H$ and to study the $x_{B}$-dependence
of the $t$-slope of the pure DVCS cross section and to study nucleon
tomography. Projections on the achievable accuracies and preliminary
results of pilot measurements will be presented.
\end{abstract}
\maketitle

\section{Introduction}

\begin{figure}
\begin{minipage}[t]{0.45\columnwidth}%
\includegraphics[width=1\columnwidth]{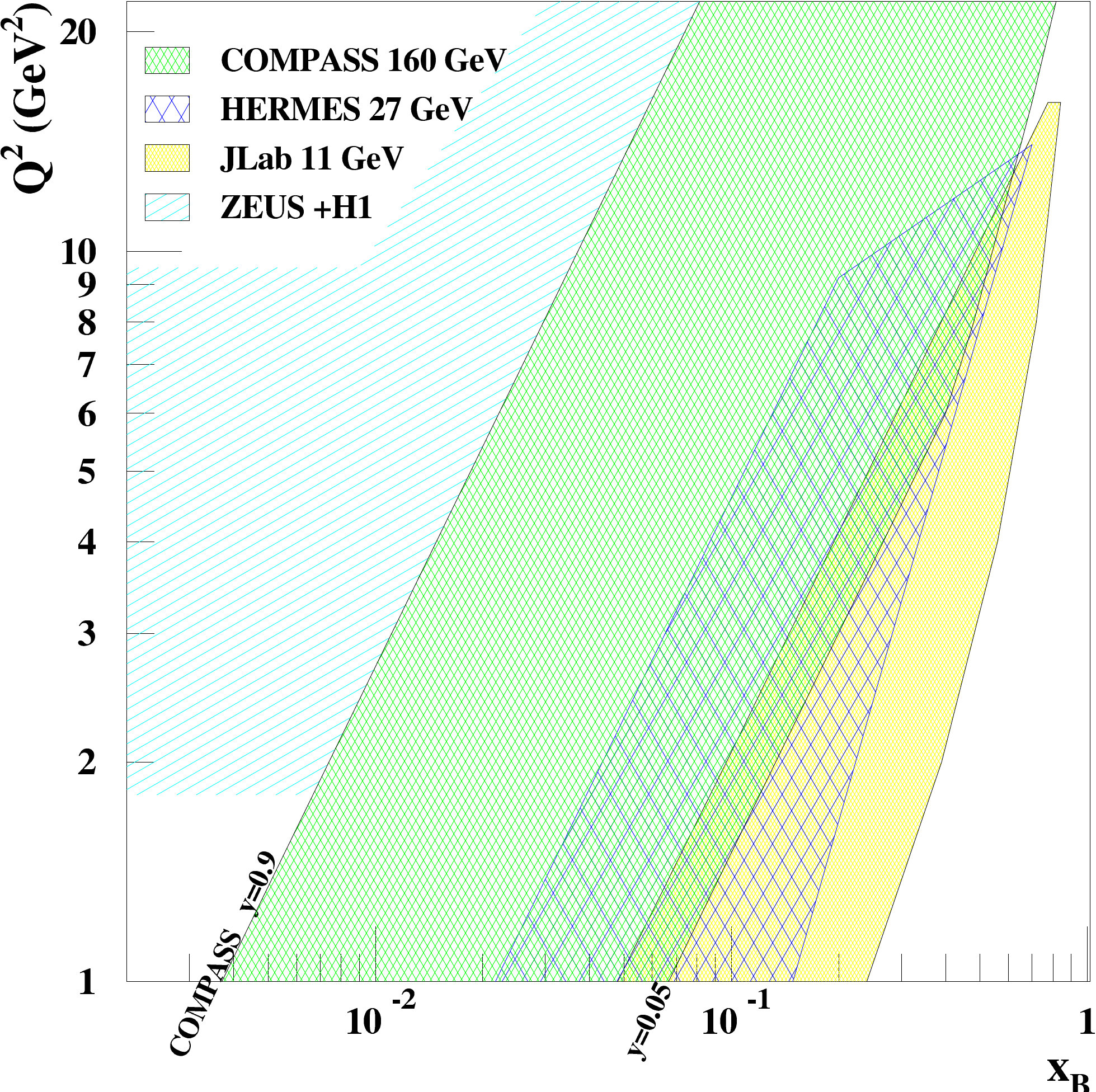}%
\end{minipage}
\begin{minipage}[t]{0.55\columnwidth}%
\includegraphics[width=1\columnwidth]{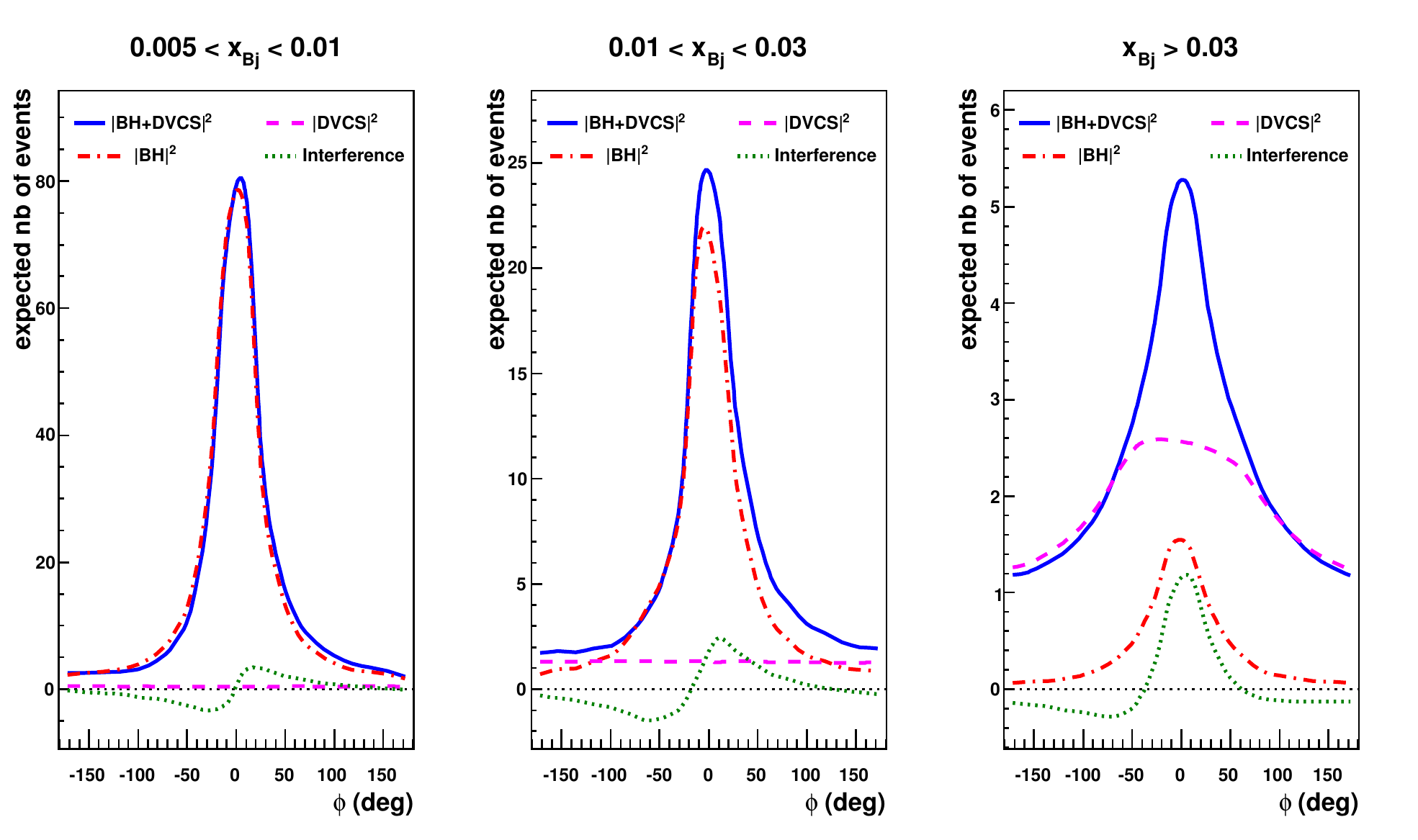}%
\end{minipage}\caption{Left: kinematic domains for measurements of hard exclusive processes
shown for the COMPASS (green area enclosed by the lines $y=0.9$ and
$y=0.05$), HERMES and JLab fixed-target experiments, and the HERA
collider experiments H1 and ZEUS.\protect \\
Right: MonteCarlo simulation of the exclusive muonproduction of single
hard photons. The plots show the $\phi$ angle distribution of reconstructed
events for three bins in $x_{B}$ and for $Q^{2}>1$~GeV$^{2}$.
}
\label{Flo:kine}
\end{figure}

Generalised Parton Distributions (GPDs)~\cite{mueller,ji1,ji2,rad1,rad2}
contain the currently most complete information on the nucleon structure.
A GPD can be considered as a momentum-dissected form-factor providing
information on the transverse localisation of a parton as function
of the fraction it carries of the nucleon's longitudinal momentum.
Hence they embody both the form factors observed in elastic scattering
and the parton distribution functions measured in deeply inelastic
scattering. GPDs provide a sort of ``3D picture'' of the nucleon
often referred as ``nucleon tomography''. Moreover it has been shown~\cite{ji1}
that the total angular momentum of partons can be accessed via the
second moment of the sum of the GPDs H and E. The study of exclusive
reactions like Deeply Virtual Compton Scattering (DVCS) and Deeply
Virtual Meson Production (DVMP) is one of the most promising ways
to experimentally constrain the GPDs. Measurements of those processes
have been performed or are planned at JLab~\cite{jlab1,jlab2,jlab3},
DESY (HERMES~\cite{hermes1,hermes2}, H1~\cite{H1_1,H1_2} and ZEUS~\cite{zeus})
and CERN (COMPASS-II)~\cite{proposal}. COMPASS-II
will cover the kinematics domain from $x_{B}\sim5\times10^{-3}$ to
about $0.1$, which cannot be explored by any other existing or planned
facility in the near future (see figure~\ref{Flo:kine}).

COMPASS uses the high-energy muon beam delivered by the M2 secondary
beam line of the SPS accelerator complex at CERN. The beam, produced
by in-flight pion decays, is naturally polarized in the longitudinal
direction (with a typical value of $\sim$80\% at 160~GeV) with an
opposite helicity sign for the two charges. The experimental apparatus~\cite{compass}
is being upgraded with a new electromagnetic calorimeter (ECAL0) and
a 4~m-long recoil proton detector (CAMERA) surrounding a 2.5 m-long
liquid-hydrogen target. The ECAL0 will extend the angular acceptance
for exclusive photons and increase the accessible $x_{B}$ domain
to substantially higher values compared to the present set-up. The
CAMERA detector on the other end will ensure the exclusivity of DVCS
and DVMP by detecting the recoil protons.

\section{Measurement of DVCS at COMPASS}

DVCS is presently the theoretically cleanest approach to GPDs, because
effects of next-to-leading order and subleading twist are under theoretical
control~\cite{belitsky}. In DVCS, a virtual photon emitted by the
incoming lepton scatters off a nucleon and emerges as a real photon
in the final state, while the target nucleon recoils without being
destroyed. DVCS competes with the Bethe-Heitler (BH) process, which
is elastic lepton-nucleon scattering with a hard photon emitted by
either the incoming or outgoing lepton. The two processes interfere
at the level of amplitudes and the differential cross-section for
hard exclusive leptoproduction of a single real photon off an unpolarised
proton target can be written as
\begin{equation}
\frac{\mathrm{d^{4}\sigma(\mu p}\to\mu p\gamma)}{\mathrm{d}x_{B}\mathrm{d}Q^{2}\mathrm{d}|t|\mathrm{d}\phi}\ =\ \mathrm{d}\sigma^{BH}\ +\ (\mathrm{d}\sigma_{unpol}^{DVCS}+P_{\mu}\mathrm{d}\sigma_{pol}^{DVCS})\ +\ e_{\mu}(\mathrm{Re}\, I+P_{\mu}\mathrm{Im}\, I)\label{eq:excl_photon_xsec}
\end{equation}
where $\phi$ is the angle between the lepton scattering plane and
the photon production plane, $e_{\mu}$ and $P_{\mu}$ are, respectively,
the charge and the polarization of the lepton beam, and $I$ represents
the DVCS~-~BH interference term. The broad kinematics domain covered
in COMPASS allows to explore regions where either the BH or the DVCS
processes dominate (see figure~\ref{Flo:kine}).

At small $x_{B}$, the almost pure BH sample will provide a precise
reference yield to accurately control the detector acceptance and
the luminosity measurement. At large $x_{B}$, where the DVCS contribution
becomes dominant, the known BH contribution can be subtracted and
the pure DVCS cross-section studied as function of $x_{B}$ and $t$.
In the intermediate region, the DVCS contribution is ``boosted''
by the BH process through the interference term. It has to be noted
that the $\phi$-dependence of the DVCS cross-section in figure~\ref{Flo:kine}
is not flat at large $x_{B}$ due to acceptance effects in the set-up
used for this simulation, which did not include the large-angle calorimeter
ECAL0.

COMPASS is presently the only facility able to perform measurements
with both beam charges of opposite helicities. This allows to isolate
various terms in equation~(\ref{eq:excl_photon_xsec}) by summing or
subtracting measurements obtained with both beam charges.

\subsection{$t$-slope of the pure DVCS cross-section to study nucleon tomography}

The unpolarized DVCS cross-section $\frac{\mathrm{d}\sigma}{\mathrm{d}t}\propto\exp(-B(x_{B})|t|)$
can be measured by considering the the the \textbf{sum} ($\mathcal{S}$)
of cross-sections with opposite beam \textbf{charge} ($C$) and \textbf{spin}
($S$) and for \textbf{unpolarized} ($U$) protons ($\mathcal{\mathcal{S}}_{CS,U}$),
integrated over $\phi$ and after subtraction of the known BH contribution.
At small $x_{B}$, 
the $t$-slope parameter $B(x_{B})$ is related to the total transverse
size $r_{\perp}$of the nucleon via the relation $<r_{\perp}^{2}(x_{B})>\approx2\cdot B(x_{B})$.
In the simple ansatz $B(x_{B})=B_{0}+2\alpha'\log(\frac{x_{0}}{x_{B}})$
the expected decrease of the nucleon size with increasing $x_{B}$
is described by the parameter $\alpha'$. Data on $B(x_{B})$ have
only been provided by the HERA collider experiments in the low $x_{B}$
range from $10^{-4}$ to $0.01$, and no significant evolution with
$x_{B}$ was observed; a first value of the transverse proton radius
$<r_{\perp}^{2}>=0.65\pm0.02$~fm has been determined using H1 data~\cite{H1_1,H1_2}.

The study of the $x_{B}$ and $t$-dependence of the DVCS cross-section
will allow us to draw conclusions on the $x_{B}$ evolution of the
transverse size of the nucleon (often referred as ``Nucleon Tomography'')
in the so far unmeasured region $0.01<x_{B}<0.1$ accessible to COMPASS.
The corresponding projected statistical accuracy is shown in figure~\ref{Flo:MC_results};
values of $\alpha'$ down to 0.125 (corresponding to half of the value
for Pomeron exchange in soft scattering processes) can be determined
with an accuracy better than $2.5\,\sigma$ with the upgraded set-up
that includes the large-angle calorimeter ECAL0.

\subsection{Determination of the Compton Form Factor $\mathcal{H}$}

The analysis of the $\phi$-dependence of the \textbf{difference}
($\mathcal{D}$) of cross-sections with opposite beam \textbf{charge}
($C$) and \textbf{spin} ($S$) and for \textbf{unpolarized} ($U$)
protons ($\mathcal{D}_{CS,U}$) will provide the two leading twist-2
expansion coefficients $c_{0}^{I}$ and $c_{1}^{I}$~\cite{proposal} that,
in the kinematics domain of COMPASS, are mainly related to the real
part of the CFF $\mathcal{H}$. This quantity was found to be positive
at H1 and ZEUS and negative at HERMES and JLab; hence, the COMPASS
kinematic domain is expected to provide the node of this amplitude,
which is an essential input for any global fit analysis.

\begin{figure}
\begin{minipage}[t]{0.49\columnwidth}%
\includegraphics[width=1\columnwidth]{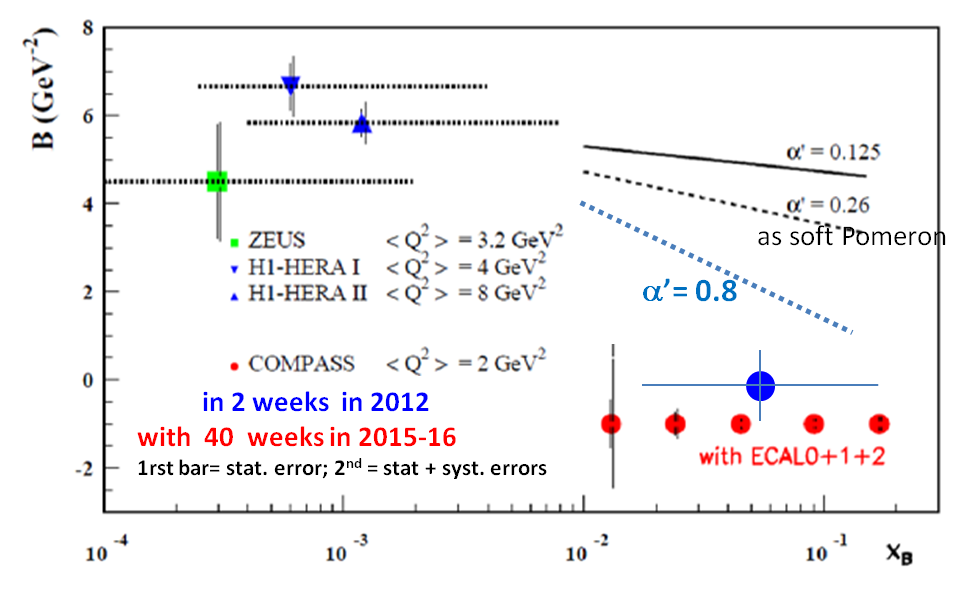}%
\end{minipage}\hfill%
\begin{minipage}[t]{0.49\columnwidth}%
\includegraphics[width=1\columnwidth]{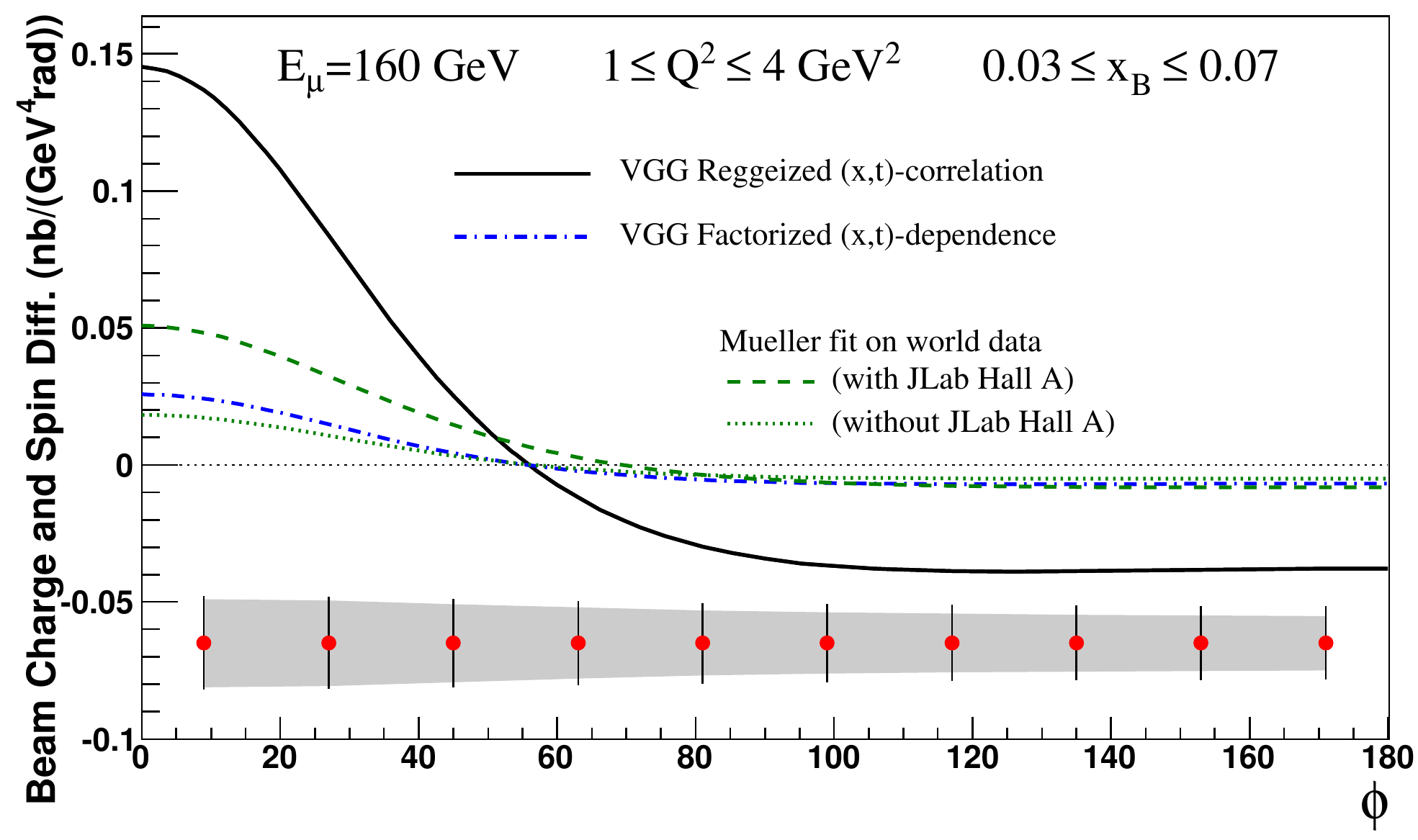}%
\end{minipage}

\caption{Left: projections for the measurement of the $t$-slope parameter
$B(x_{B})$ for various bins in $x_{B}$. The bars on the red data
points show the expected statistical (left bar) and total (right bar)
errors assuming a running time of 280 days with a 2.5~m long liquid
H$_{2}$ target. The solid and
dashed lines show the results of two parametrizations corresponding
to $\alpha'=0.125$~GeV$^{-2}$ and $\alpha'=0.26$~GeV$^{-2}$. The blue 
round point shows the expected statistical accuracy assuming two weeks
of data taking in 2012.
\protect \\
Right: expected statistical (error bars) and systematic (grey band)
accuracy for the measurement of the $\phi$-dependence of $\mathcal{D}_{CS,U}$
for $1\le Q^{2}\le4$~GeV$^{2}$ and $0.03\le x_{B}\le0.07$. A running
time of 280 days with a 2.5~m long liquid H$_{2}$ target has been
assumed for the estimate. The solid and dash-dotted curves correspond
to different variants of the ``VGG'' model~\cite{VGG}, while the
other two curves show predictions based on fits to existing HERA,
HERMES and JLab data~\cite{kum1,kum2}.}
\label{Flo:MC_results}
\end{figure}

The expected statistical (error bars) and systematic (grey band) accuracy
for the measurement of the $\phi$-dependence of $\mathcal{D}_{CS,U}$
in a particular $(x_{B},Q^{2})$ bin is shown in figure~\ref{Flo:MC_results}.
Two of the curves are calculated using the ``VGG'' GPD model~\cite{VGG}
using either a ``reggeized'' parametrization of the correlated $x,t$
dependence or a ``factorized'' $x,t$ dependence of the GPDs; here
$x$ represents the average between the initial and final longitudinal
momentum fractions of the nucleon, carried by the parton throughout
the process. The other two curves are the result of a fitting procedure~\cite{kum1,kum2},
including next-to-next-to leading order (NNLO) corrections, which
was developed and successfully applied to describe DVCS observables
from both very small $x_{B}$ (HERA) and large $x_{B}$ (HERMES and
JLab).

In a similar manner, the analysis of the $\phi$-dependence of $\mathcal{\mathcal{S}}_{CS,U}$
will provide information on the imaginary part of the CFF $\mathcal{H}$.

\bibliographystyle{aipproc}   

\bibliography{ferrero_diffraction2012}

\end{document}